\documentclass[%
 aip,
 amsmath,amssymb,
 reprint,%
]{revtex4-1}

\usepackage{graphicx}
\usepackage{dcolumn}
\usepackage{bm}

\usepackage[utf8]{inputenc}
\usepackage[T1]{fontenc}
\usepackage{mathptmx}
\usepackage{etoolbox}
\usepackage[,]{siunitx}
\usepackage{makecell}

\makeatletter
\def\@email#1#2{%
 \endgroup
 \patchcmd{\titleblock@produce}
  {\frontmatter@RRAPformat}
  {\frontmatter@RRAPformat{\produce@RRAP{*#1\href{mailto:#2}{#2}}}\frontmatter@RRAPformat}
  {}{}
}%
\makeatother
\begin{document}

\preprint{AIP/123-QED}

\title{High-sensitivity silicon nitride microring resonator opto-fluidic sensor}

\author{D.O. Armstrong}
\affiliation{ 
School of Engineering, Cardiff University, Queen's Buildings, The Parade, Cardiff, UK, CF24 3AA}
\affiliation{Translational Research Hub, Maindy Road, Cardiff, CF24 4HQ, UK}

\author{S. Ibrahim}
\affiliation{ 
School of Engineering, Cardiff University, Queen's Buildings, The Parade, Cardiff, UK, CF24 3AA}
\affiliation{Translational Research Hub, Maindy Road, Cardiff, CF24 4HQ, UK}

\author{S. Naserikarimvand}
\affiliation{ 
School of Biological and Marine Sciences, Faculty of Science and Engineering, University of Plymouth, Drake Circus, Plymouth, PL4 8AA, UK}

\author{S. Whelan}
\affiliation{ 
Faculty of Science and Engineering, University of Plymouth, Drake Circus, Plymouth, PL4 8AA, UK}

\author{O.J. Guy}
\affiliation{ 
Department of Chemistry, School of Engineering and Applied Sciences, Faculty of Science and Engineering, Swansea University, Swansea, SA2 8PP, UK}

\author{A.J. Bennett}
\affiliation{ 
School of Engineering, Cardiff University, Queen's Buildings, The Parade, Cardiff, UK, CF24 3AA}
\affiliation{Translational Research Hub, Maindy Road, Cardiff, CF24 4HQ, UK}

\author{J.P. Hadden}
\affiliation{ 
School of Engineering, Cardiff University, Queen's Buildings, The Parade, Cardiff, UK, CF24 3AA}
\affiliation{Translational Research Hub, Maindy Road, Cardiff, CF24 4HQ, UK}

\date{\today}
             
\begin{abstract}
Photonic integrated circuit devices can be used as refractometric opto-fluidic sensors to detect the presence of analytes in solution at low concentrations. In this work, we investigate the refractive index sensitivity of silicon nitride microring resonator based photonic integrated circuit fluidic sensors. The performance of a foundry fabricated sensor is measured over the C-band in the presence of liquid samples achieving a mean sensitivity of 579 nanometres per refractive index unit. This demonstration of a scalable, high-sensitivity opto-fluidic sensor, compatible with recognition marker surface functionalisation, opens the way to applications in environmental and bio-sensing.
\end{abstract}
\maketitle

\section{\label{sec:introduction}Introduction}

In recent years there has been an increase in the demand for medical diagnostic tools which can be used outside of laboratory settings, in clinics, or directly by patients. The increased adoption of self-monitoring apparatus for personal healthcare have accelerated the research and development of low-cost, highly-deployable biochemical sensors, facilitating the lab-on-chip paradigm \cite{Estevez2012}. One application of miniaturised biochemical sensors is in the field of environmental monitoring. Contaminants in water supplies—from pathogenic microorganisms and viruses to toxic chemicals—pose significant risks to both public health and the environment. Yet, current monitoring approaches remain slow, expensive, and heavily reliant on centralised laboratory facilities, creating a bottleneck in environmental analysis. By leveraging photonic integrated circuits (PICs) with tailored surface chemistries via functionalisation, solutions to this bottleneck can offer highly sensitive, real-time analysis without the need for lengthy sample processing. This enabling technology will not only provide earlier intervention and safer water systems, but also demonstrate the transformative potential of integrated photonic sensors for widespread adoption in environmental, agricultural, and healthcare applications.

Sensors based on optical microring resonators (MRRs) are among the leading chip-scale photonic sensors that have been demonstrated. The working principle of these devices is that optical resonances of the MRRs are affected by the interaction of the resonant modes' evanescent field with the components of a liquid solution in which the MRR is immersed. Surface-functionalisation of the MRR with recognition markers which bind to specific targets within the solution then allows transduction of the target capture to detectable shifts in the optical resonant wavelength. In order to maximise the sensitivity of MRR biosensing devices therefore, the device should be engineered such that the optical resonances are as sensitive as possible to changes in the surrounding refractive index due to the presence of a particular analyte of interest (AOI). 

MRR sensors have been demonstrated featuring optical resonances with moderate-to-high quality (Q) factors of \SIrange[parse-numbers=false,range-units=single]{10,000}{100,000}, offering refractive index sensitivities of hundreds of nanometres of resonance shift per refractive index unit (RIU), low detection limits (\SIrange[parse-numbers=false,range-units=single]{10^{-7}}{10^{-4}}{RIU}) and compact footprint of up to hundreds of square microns \cite{DeVos2007,Ciminelli2013,Sedlmeir2014}. MRRs are also favourable for their ability to be fabricated in dense arrays, allowing for multiplexed analysis. Various different implementations of MRR-based fluidic sensors have been demonstrated in the literature. A summary of recently demonstrated MRR-based fluidic sensors fabricated on SiN and silicon-on-insulator (SOI) platforms can be found in Table \ref{tab:1}. Of particular note is Chen, et al's demonstration, with a sensitivity of \SI{5752.5}{nm/RIU}, consisting of a feedback-coupled microring resonator device which includes a more complicated MRR geometry. To the author's knowledge is the highest reported sensitivity in the literature.

\begin{table}[b]
\caption{Examples of MRR-based fluidic sensors in the literature.} \label{tab:1}
\begin{center}
\begin{tabular}{|c|c|c|c|c|}
 \hline
\makecell{Waveguide \\type} & \makecell{Sensitivity \\ (nm/RIU)} & Q-factor & \makecell{Mode/\\Band} & Reference \\ 
 \hline 
 SiN Strip & 579 & 23,000 & TE/C & This work\\ 
 \hline
 SiN Strip & 179.7 & 24,000 & TE/C & Bryan, et al. \cite{Bryan2023} \\ 
 \hline
 SiN Strip & 200 & - & TM/O & Castelló-Pedrero, et al. \cite{Lucia2023} \\ 
 \hline
 SOI Slot & 433.3 & 4300 & TE/C & Liu, et al. \cite{Liu2021} \\
 \hline
 SOI Porous Slab & 380 & 10,000 & TE/C & Rodriguez et al. \cite{Rodriguez2015} \\ 
 \hline
 SOI Strip & 5752.5 & - & TE/C & Chen, et al. \cite{Chen2025} \\ 
 \hline
\end{tabular}
\end{center}
\end{table}

\begin{figure*}[ht]
    \centering
    \includegraphics[scale = 0.28]{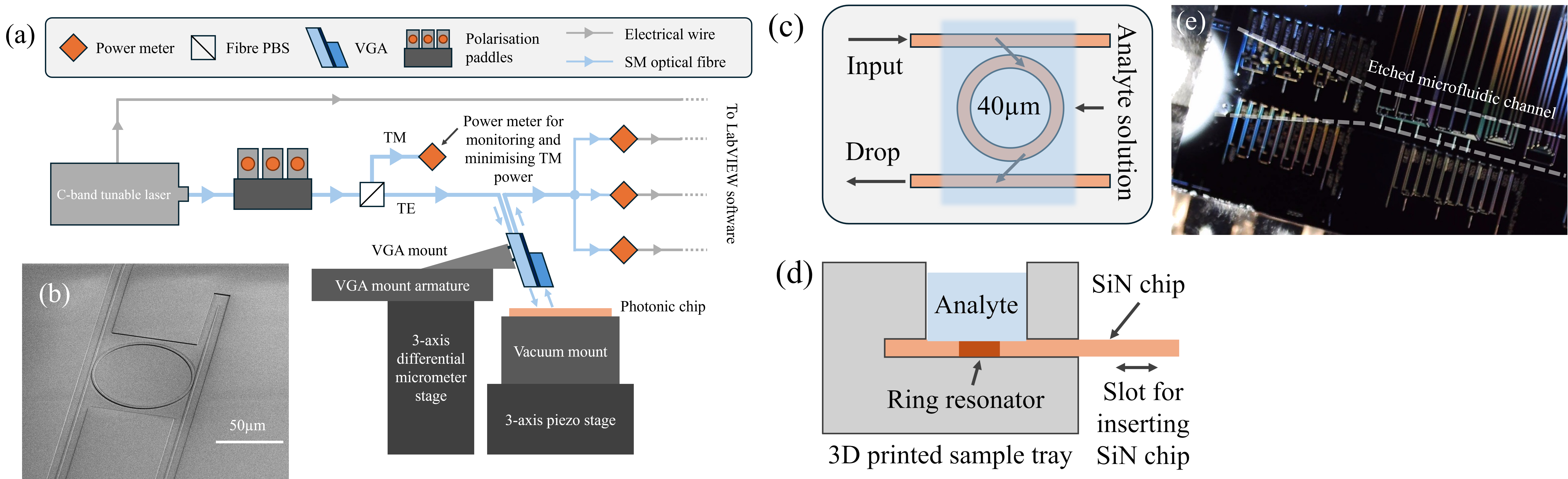}
    \caption{Experimental setup. (a) Schematic of the experimental setup used to characterise our SiN MRRs. (b) SEM micrograph of a fabricated \SI{40}{\micro m} radius MRR.
    (c) Schematic of the SiN MRR immersed in analyte. (d) Schematic cross section of the SiN sample tray showing how analyte solutions are  delivered to the sensing microring. (e) Camera image of the SiN photonic chip with etched channel shown.}
    \label{fig:experimental_setup}
\end{figure*} 

Intuitively, high Q-factor microcavity sensors offer increased precision through sharper resonances, however these devices typically have reduced interaction between the cavity mode and the AOI due to their more confined field. The narrow resonance bandwidth, which requires a strict resonance spectral alignment, is also sensitive to temperature fluctuation and other environmental factors which may cause undesired resonance shifts, thus compromising reliability. Therefore, devices with lower Q-factor but more interaction with the AOI may provide higher refractive index sensitivity.

In this work, we investigate the sensitivity of a foundry fabricated silicon nitride (SiN) microring resonator-based photonic integrated circuit fluidic sensor. SiN is a material platform which is especially attractive for PIC sensors due to its wide transparency window (\SI{300}{nm} to beyond \SI{8000}{nm}) low-loss waveguides (\SI{<0.1}{dB/cm}) \cite{Guo2023}, comparable to its silicon-on-insulator counterparts and low thermo-optic coefficient (\SI{2.45e-5}{RIU/K})\cite{Arbabi2013}. It offers wide refractive index tuneability through stoichiometry adjustment allowing great flexibility in optical waveguide design. This means that waveguides and modal characteristics can tuned to the need of the PIC sensor; enabling low-index-contrast device designs for fibre-waveguide mode matching or alternatively high-index-contrast components for low-loss propagation. SiN photonic fabrication is also CMOS compatible, allowing scalable fabrication and integration with integrated electronics\cite{Payne2022}. The sensor’s optical response is investigated over the C-band from \SI{1528}{nm} to \SI{1568}{nm}  in the presence of liquid analytes, demonstrating a mean sensitivity of 579 nanometres per refractive index unit.

\section{\label{sec:theoryofsensors} Microring resonator refractometric sensors}

Refractometric measurements for a MRR in the presence of a certain AOI involves the tracking of shifts to the resonant modes in wavelength in response to a change in the effective refractive index $n_{eff}$ which they experience within the sensing region. This shift can be measured by scanning a laser across the resonance and measuring its position in wavelength through transmission spectroscopy. 

The MRR's resonant modes are related to the microring resonance condition through 

\begin{equation} \label{eq:ResonantWavelength}
    \lambda_{res} = 2\pi n_{eff} R / m
\end{equation}

where $\lambda_{res}$ is the resonant wavelength,  $R$ is the effective radius of the mode in the MRR, and $m$ is the azimuthal mode order \cite{Estevez2012}. Thus a change in the effective index will cause a detectable change in the resonant wavelengths. 

We also define the Q-factor, which relates to the average number of round trips the light makes around the ring before it is lost to internal losses, the sensitivity, $S$, which is determined by the extent of the overlap between the AOI and the evanescent field and the limit of detection (LOD), which refers to the smallest detectable change in the surrounding refractive index. These are expressed as 

\begin{equation} \label{eq:Q-factor}
    Q = \frac{\lambda_{res}}{\Gamma},
\end{equation}

where $\Gamma$ is the full-width half-maximum of the resonance,

\begin{equation} \label{eq:Sensitivity}
    S = \frac{\Delta\lambda_{res}}{\Delta n_{clad}},
\end{equation}

and 

\begin{equation} \label{eq:LOD}
    LOD = \frac{\lambda_{res}}{QS}.
\end{equation}

\section{\label{sec:results} Experimental Results}

\subsection*{\label{subsec:ringcharacterisation} Microring resonator characterisation}

PIC microring resonator sensor chips were fabricated through the \SI{300}{nm} SiN multi-project-wafer service at the CORNERSTONE foundry\cite{Littlejohns2020}. Devices are patterned in a \SI{300}{nm} epilayer of Si$_3$N$_4$ grown by low-pressure chemical vapour deposition (LPCVD) on a \SI{3}{\micro m} buried oxide layer of thermal silica (SiO$_2$) on a crystalline silicon substrate. Devices are then clad in a \SI{2}{\micro m} thick layer of SiO$_2$ deposited by plasma-enhanced chemical vapour deposition (PECVD). Subsequently, a channel is etched in the vicinity of the microring resonator sensors allowing controlled exposure to analytes. 

The characterised device is a \SI{40}{\micro m} radius MRR in the `add-drop' configuration \cite{Bogaerts2012}.  This is expected to provide high refractive index sensitivity based on 2D finite-difference time-domain (FDTD) modelling presented in section \ref{subsec:claddingsimulations}. To characterise the devices, a C-band tunable continuous-wave laser with a linewidth of \SI{100}{\kilo Hz} was coupled into and out of the bus waveguide with a 1D grating coupler via a fibre array with piezo-controlled positioning. TE input polarization was pre-selected using a paddle-style polarisation controller. A schematic for the characterisation setup is shown in Figure \ref{fig:experimental_setup}(a), while an SEM micrograph of a fabricated MRR is shown in Figure \ref{fig:experimental_setup}(b).

\begin{figure}[ht!]
    \centering
    \includegraphics[scale = 0.22]{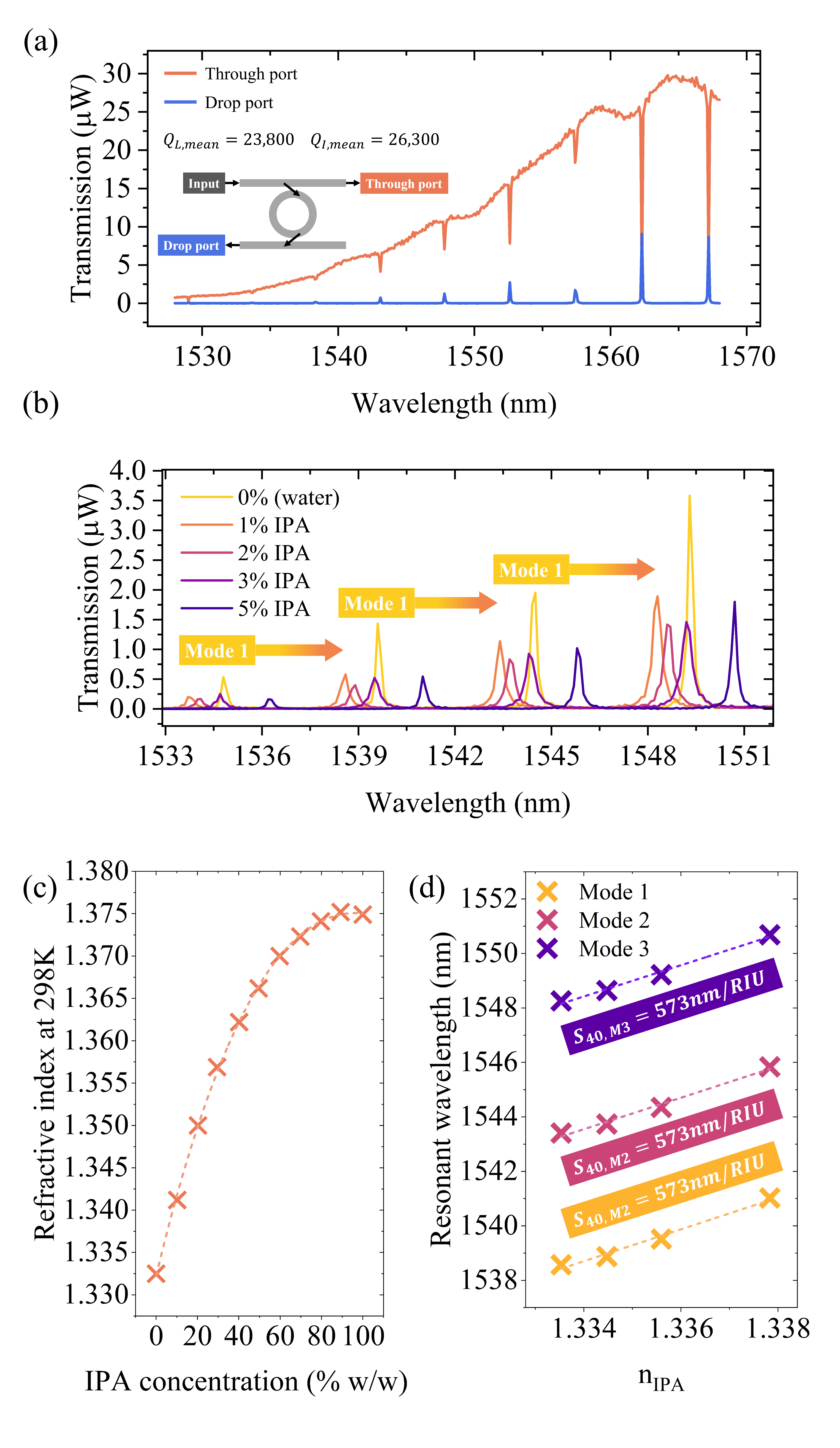}
    \caption{Optical characterisation. (a) Transmission spectra of a \SI{40}{\micro m} radius SiN microring resonator with a mean FSR of \SI{4.83}{nm}. The orange trace corresponds to the `though port' and the blue trace corresponds to the `drop port'. (b) Experimental characterisation of a \SI{40}{\micro m} SiN microring in the presence of varying concentrations of IPA. The predicted progression of the resonance modes is shown. (c) IPA solution refractive index as a function of $\%$ (w/w) concentration. (d) Resonance wavelength shift as a function of IPA refractive index, from which we calculate the sensor's sensitivity.}
    \label{fig:experimental_characterisation}
\end{figure}

Performing optical characterisation of our \SI{40}{\micro\meter} radius microring revealed a series of resonances with a mean FSR of \SI{4.83}{nm} as shown in Figure \ref{fig:experimental_characterisation}(a). Greater optical intensity is seen at the `red' end of the spectrum, where the grating couplers are more efficient. Between resonances from \SI{1558}{nm} onwards there is a low-visibility sinusoidal oscillation resulting from a weak Fabry-P\'erot cavity formed by back reflections from the grating couplers. Analysis of the microring's Q-factor reveals a mean loaded Q-factor, $Q_{L,mean}$, of 23,800 and a mean intrinsic Q-factor, Q$_{I,mean}$ of 26,300 in air.

\subsection*{\label{subsec:ipaconcentration} Experimental opto-fluidic sensor sensitivity}

Resonance shift measurements were taken in the presence of varying concentrations of isopropyl alcohol (IPA) solution, in order to characterise the overall resonator sensitivity and limit of detection (LOD). IPA solutions were prepared prior to our measurements and consisted of \si{1}{\%}, \SI{2}{\%}, \SI{3}{\%}, \SI{4}{\%}, and \SI{5}{\%} solutions. For each IPA solution, water and IPA were mixed volumetrically to achieve the desired concentration. The uncertainty values are based on the reported percentage error for the particular Eppendorf pipette we used. For our purposes we used a 30-300$\mu$L pipette for measuring out the IPA and a 500-5000$\mu$L pipette for measuring the deionised water. The reported systematic error for both of these pipettes is $\pm$0.6$\%$. Between each measurement, the sample tray was evacuated and the new IPA concentration was added to the sample tray ready for the next measurement.

Ambient temperature is known to affect the optical response and sensitivity of photonic sensors. This is a direct result of the thermo-optic effect, whereby the refractive index of the waveguiding material changes with temperature. In order to ensure our biosensors thermal stability during liquid analyte measurements, we placed the sample on top of a heating block connected to a thermoelectric controller (TEC) to keep the bulk temperature of the sample constant. We also measured the temperature-dependent variability of our MRR's resonant modes, finding that over a \SI{10}{K} temperature range, the resonance is stable and shifts by only \SI{13(2)}{pm/K}.

A channel etched  in the vicinity of the microring resonator sensor enabled the sensing devices to be exposed to fluid samples. The photonic chip was packaged in a simple 3D printed sample tray defining an open reservoir. A schematic of this tray can be seen in Figure \ref{fig:experimental_setup}(c) and (d) while a camera image of the sample with the etched channel is in Figure  \ref{fig:experimental_setup}(e). 

As shown in Figure \ref{fig:experimental_characterisation}(c), the resonance wavelengths of the microring show a steady redshift as the IPA concentration is increased. For modes labelled 1-3 we calculate sensitivities of \SI{585(3)}{nm/RIU}, \SI{578(3)}{nm/RIU}, and \SI{573(3)}{nm/RIU} respectively, with a mean sensitivity of \SI{579}{nm/RIU}. This value exceeds the predicted sensitivity obtained via simulation, which we attribute to the underestimation of the evanescent field interaction with the sensing region due to the use of the variational FDTD solver. Using Equation \ref{eq:Sensitivity}, as well as the calculated Q-factor of 23,000 we calculate the LoD of the resonant mode at \SI{1539.6}{nm} to be \SI{2.55e{-4}}{\%}. This indicates that the smallest concentration of IPA that would produce a detectable resonance shift is on the order of \SI{1e-4}{\%}.

Currently, the performance of our sensor is limited by noise and wavelength calibration. Noise on the measurements of our uncalibrated resonance wavelength limits the efficacy to which we can determine shifts in the resonance as liquid analytes are introduced. Assuming that we have narrowband resonances, minimal dispersion and that the minimum detectable resonance shift, $\delta\lambda_{min}$, is comparable to the resonance FWHM, $\Gamma$, then the smallest physical change to in the the cladding index, which produces a resonance shift we can reliably distinguish from background noise, $\Delta n_{c,min}$, is given by 

\begin{equation} \label{eq:5}
    \Delta n_{c,min} \approx \frac{\lambda}{QS \cdot SNR}, 
\end{equation}

where the $SNR$ is the signal-to-noise-ratio of the resonance. Equation \ref{eq:5} tells us that the smallest detectable change to the resonance wavelength is inversely proportional to the SNR. Therefore, we may improve the limit of detection of our sensor by increasing the resonance amplitude through careful device design.

\section{\label{subsec:claddingsimulations} Modelling silicon nitride microring resonators}

\begin{figure}[ht]
    \centering
    \includegraphics[scale = 0.24]{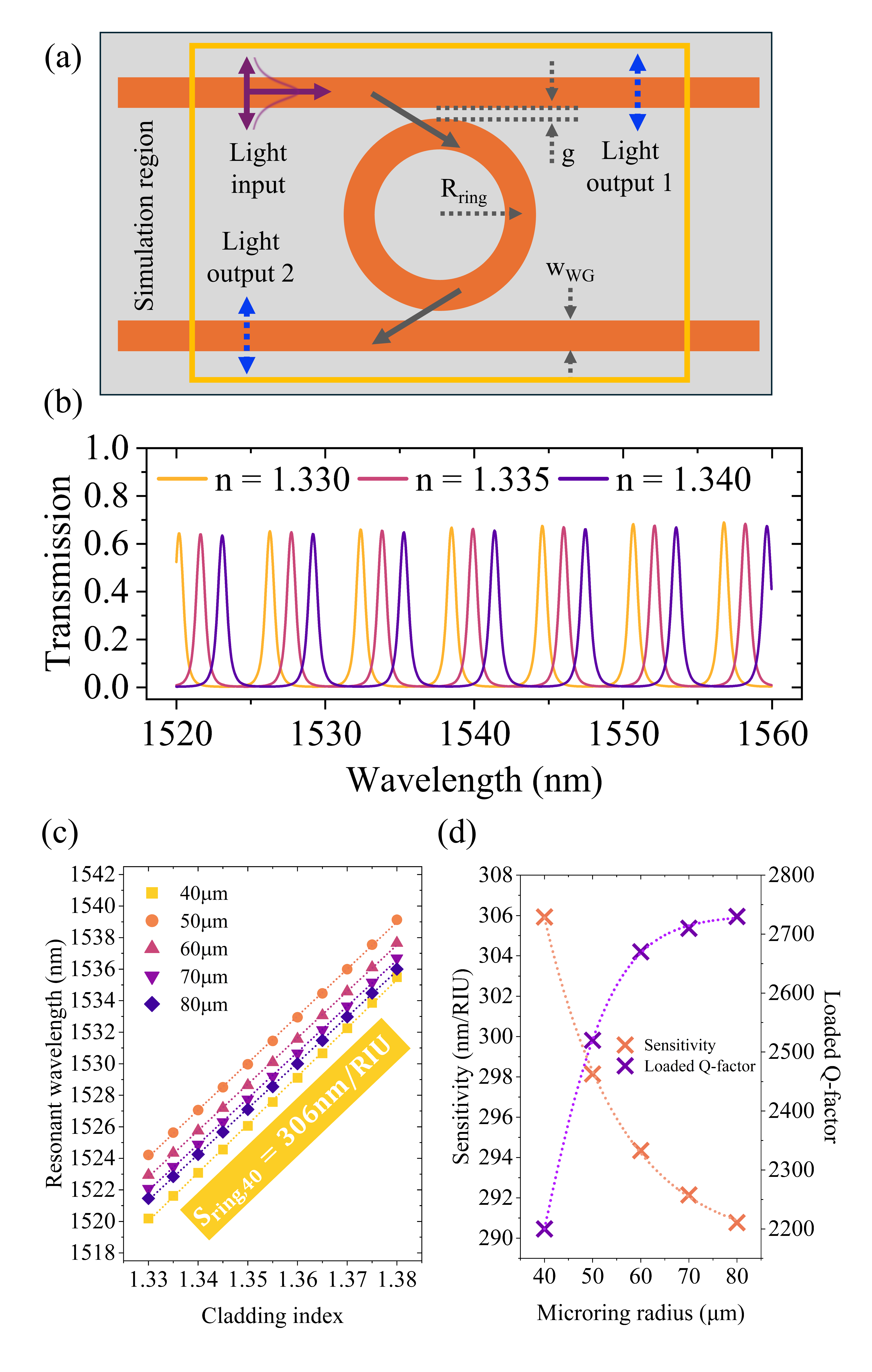}
    \caption{Simulation results of a \SI{40}{\micro m} SiN microring resonator. (a) Schematic showing the simulation setup. (b) Transmission spectra of the \SI{40}{\micro m} microring showing the first three cladding refractive indices. (c) Resonance wavelength shift as a function of the cladding refractive index for different microring radii. We chose the optical mode at $\approx$\SI{1526}{nm} with a cladding index of 1.330 as the starting point to measure subsequent shifts in the resonance wavelength. (d) Microring sensitivity and loaded Q-factor as a function of microring radius.}
    \label{fig:simulations}
\end{figure} 

We perform 2D finite-difference time-domain (FDTD) simulations of SiN MRRs using Lumerical's commercial varFDTD solver to further understand the MRR's use as a sensor. Transmission spectra of the SiN MRR as were acquired as the microring's radius was varied from \SI{40}{\micro m} to \SI{80}{\micro m} and the cladding refractive index from 1.33 to 1.38. The SiN waveguide and ring resonator dimensions were set to match the geometry of our fabricated devices, with waveguide dimensions of \SI{1.2}{\micro m} by \SI{0.3}{\micro m} and a coupling gap of \SI{0.35}{\micro m}. The SiN material model used in simulation was extracted from CORNERSTONE's results for the refractive index of stoichiometric SiN (n = 2.026 at \SI{1550}{nm}). The results for this analysis are presented in Figure \ref{fig:simulations}.

Figure \ref{fig:simulations}(b) shows a clear shift in the resonance wavelengths as the cladding refractive index is increased. For each radius, the resonance wavelength shift is approximately linear with cladding index over the investigated range as shown in Figure \ref{fig:simulations}(c). Calculations of the microring sensitivity values reveal that it decreases as the radius is increased as shown in Figure \ref{fig:simulations}(d). Smaller microring radii push the optical mode further into the outer edge of the microring waveguide, thereby increasing the overlap of the modal evanescent field and the surrounding sensing region. Therefore, despite the increase in optical path length, and sampling length of the analyte as the radius is increased, the results in Figure \ref{fig:simulations}(d) suggest that for single mode MRR, the bulk refractive index sensitivity has a greater dependence on the field-analyte overlap than on the increase in sampling length. 

\section{\label{conclusion}Conclusion}
We have demonstrated a SiN-on-insulator microring resonator for use as a refractometric sensor. The moderate Q-factor and large FSR enable us to demonstrate sensitivities exceeding other comparable sensors in a device compatible with foundry scale fabrication. The on-chip nature of these devices and the CMOS foundry compatibility allow for the rapid, low-cost fabrication of microring resonator sensors at scale, paving the way for commercial viability for this technology. Temperature stability of the microring's resonant modes is verified and the sensitivity with respect to low concentrations of analytes in solution is calculated.

In future, other photonic devices may be considered with an expressed focus on those that maximise the field overlap with the sensing region around the device. This would help us to further enhance the sensitivity. Furthermore, more sophisticated microfluidic packaging could be explored in order to better facilitate analyte delivery to the sensing devices and make the replacement of analyte solutions more convenient. Currently, the primary factor limiting our sensor's performance is the overlap of the evanescent field with the sensing region which may be improved by considering devices such as photonic crystal microring resonators or by introducing a surface chemistry treatment to facilitate sensing only in a small volume around the device. 

In order to realise a deployable water contaminant sensor, we must adopt a functionalisation scheme targetting the specific binding of waterborne pollutants (such as heavy metal ions) to the surface of our sensing MRR. Such a scheme would use surface silanisation and thiolated-aptamer binding to create functional groups that preferentially bind to metal ions dispersed in contaminated water supplies. This approach would require us to minimise the volume of water needed for a sensing measurement, decrease the time to prepare the sensor between measurements and integrate light sources and detectors on chip to create a sensor independent from bulk optics.

\begin{acknowledgments}

We wish to acknowledge the support of the EPSRC-funded CORNERSTONE National Research Facility EP/W035995/1 for providing the academic free-of-charge multi-project-wafer silicon nitride photonics fabrication capability, and Technology Readiness Intervention funding support from EPSRC grant EP/Z531066/1. We also wish to acknowledge the European Regional Development Fund (ERDF) and Welsh Government funded ASSET project (Application Specific Semiconductor Etching Technology), led by Swansea University, and project partners, Wave Photonics. Simulations made use of high-performance computing facilities provided by Advanced Research Computing @ Cardiff (ARCCA). Device processing was carried out in the cleanroom of the ERDF-funded Institute for Compound Semiconductors (ICS) at Cardiff University.

\end{acknowledgments}

\section*{Data Availability Statement}

Data supporting the findings of this study are available
in the Cardiff University Research Portal at
https://doi.org/10.17035/cardiff.29610791

\nocite{*}

\bibliography{references}

\end{document}